\documentclass[a4paper]{article}

\usepackage{INTERSPEECH2021}

\usepackage{tikz}
\usepackage{hyperref}
\usepackage{booktabs}
\usepackage{enumitem}
\usepackage{algorithm}
\usepackage[noend]{algpseudocode}
\usepackage{amsmath}
\usepackage{multirow}
\usepackage{bbm}
\usepackage{array}

\title{User-Level Differential Privacy against Attribute Inference Attack of Speech Emotion Recognition in Federated Learning}
\name{Tiantian Feng, Raghuveer Peri, Shrikanth Narayanan}
\address{
  Signal Analysis and Interpretation Lab (SAIL), University of Southern California}
\email{tiantiaf@usc.edu, rperi@usc.edu, shri@ee.usc.edu}

\newtheorem{definition}{Definition}[section]
\newtheorem{theorem}{Theorem}[section]

\begin{document}

\maketitle
\begin{abstract}

Many existing privacy-enhanced speech emotion recognition (SER) frameworks focus on perturbing the original speech data through adversarial training within a centralized machine learning setup. However, this privacy protection scheme can fail since the adversary can still access the perturbed data. In recent years, distributed learning algorithms, especially federated learning (FL), have gained popularity to protect privacy in machine learning applications. While FL provides good intuition to safeguard privacy by keeping the data on local devices, prior work has shown that privacy attacks, such as attribute inference attacks, are achievable for SER systems trained using FL. In this work, we propose to evaluate the user-level differential privacy (UDP) in mitigating the privacy leaks of the SER system in FL. UDP provides theoretical privacy guarantees with privacy parameters $\epsilon$ and $\delta$. Our results show that the UDP can effectively decrease attribute information leakage while keeping the utility of the SER system with the adversary accessing one model update. However, the efficacy of the UDP suffers when the FL system leaks more model updates to the adversary. We make the code publicly available to reproduce the results in \href{https://github.com/usc-sail/fed-ser-leakage}{https://github.com/usc-sail/fed-ser-leakage}.

\end{abstract}
\vspace{1mm}

\footnote{This paper was submitted to Insterspeech 2022 for review.}

\vspace{-3mm}
\section{Introduction}

Speech emotion recognition (SER) has found increasing applications in virtual assistants \cite{lee2020study}, health \cite{ramakrishnan2013speech, Bone2017SignalProcessingandMachine}, education \cite{li2007speech} and other emerging human-centered AI applications. SER is prone to privacy leakage issues like other speech technologies because the collected speech data can reveal sensitive information about an individual, including intent, demographic/personality traits, and health states. Federated Learning~(FL) methods attempt to address the issues of data privacy by training a model on a central server using the shared model parameters from an edge device without the need for local data \cite{mcmahan2017communication}.
However, as reported in our prior work, SER applications trained in an FL setup are still vulnerable to attribute inference attacks \cite{feng2021attribute}. In particular, we found that an adversary with access to local parameter updates can successfully infer the gender of the user (deemed as sensitive in that particular SER use case) operating the edge device. In this work, we propose to apply a recently developed user-level differential privacy~(UDP) framework \cite{wei2021user} to mitigate attribute information leakage in FL-based SER systems.

In FL algorithms, each edge device trains a local model using its own data, and the central server then aggregates the shared local model parameters. Such a training scheme ensures that local data is not shared with the central server, potentially mitigating privacy leakage. However, recent works have shown that adversaries may still perform privacy attacks, such as membership inference attacks \cite{melis2019exploiting} and reconstruction attacks \cite{zhu2020deep, geng2021towards}, by using the model parameters shared with the central server. For instance, many works have demonstrated that data reconstruction is achievable through analyzing the model updates in FL setup \cite{melis2019exploiting, zhu2020deep, geng2021towards}. We had previously demonstrated this phenomenon in FL-based SER setup \cite{feng2021attribute}. Specifically, we showed that an attribute inference attacker could successfully infer a user's gender attribute by using the model updates shared in the FL setup \cite{feng2021attribute}. A typical approach to protect privacy in FL is differential privacy (DP) \cite{wang2019beyond, wei2020federated}, of which local DP (LDP) is a prominent example \cite{wei2021user}. For instance, user-level DP, a particular LDP approach, provides privacy protections to FL by perturbing each client's shared model before uploading it to the central server. In UDP, the training process of each client satisfies the requirement of $(\epsilon, \delta)$-LDP for different privacy levels by adapting Gaussian noise with appropriate variances.

In this work, we perform an extensive exploration of this framework within the context of FL-based SER. In particular, we investigate the effect of the level of perturbation on privacy leakage and the \textit{utility} of the trained SER model. In addition, we enhance the capability of the privacy attacker by providing access to multiple model updates for each client in the FL training setup. Our experiments show that when the adversary has only access to a single model update from a client, the UDP can effectively decrease attribute information leakage (thereby mitigating privacy leakage) while retaining the utility of the SER model. However, the efficacy of this mitigation strategy drops substantially when the attacker can observe multiple model updates from the FL process.

\vspace{-1.25mm}
\section{Method}
\label{sec:method}

\begin{figure}
	\centering
	\includegraphics[width=0.8\linewidth]{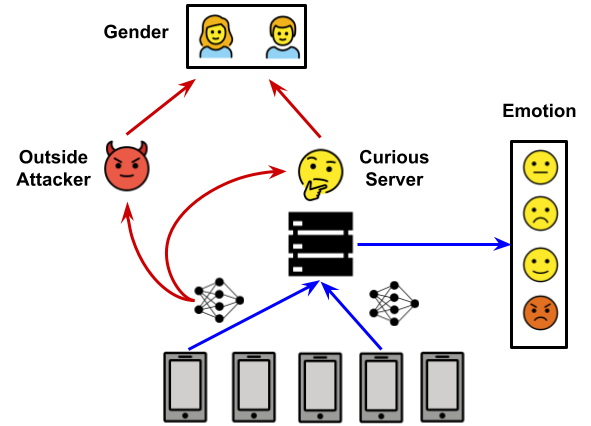}
    \caption{The figure shows the problem setup of the attribute inference attack in \cite{feng2021attribute}. Here, we define SER as the primary application, and the adversary attempts to infer the gender using the shared model updates. (Image credit: OpenMoji \cite{openmoji})}
    \label{fig:attack_problem}
    \vspace{-4mm}
\end{figure}

In this section, we first review the attacking framework we proposed in \cite{feng2021attribute}. We then summarise the proposed UDP algorithm used in this work. To facilitate readability, we summarize the notations adopted in this paper in \autoref{tab:notation}. \autoref{fig:attack_problem} shows the attack problem setup we apply in this work. Specifically, the \textit{primary task} is SER, models for which are trained using the FL framework. In contrast, in the \textit{adversarial task} the attacker attempts to predict the client's gender label (deemed sensitive in this exemplary scenario). We follow a setup in which we have a private-labeled data set $\mathcal{D}^{p}$ from a number of clients, where each client has a feature set $\mathbf X$ and an emotion label set $\mathbf y$. Each client is also associated with a gender label $z$. In this attack, the adversary tries to infer the sensitive attribute $z_{k}$ of the $k^\text{th}$ client using global model $\mathbf{\theta^{t}}$ and its local model $\mathbf{\theta_{k}^{t+1}}$.

\begin{table}
    \centering
    \caption{Notation used in this paper.}
    \vspace{-2mm}
    \begin{tabular}{p{0.5cm}p{0.5cm}}
        \toprule

        \multicolumn{1}{l}{$\mathcal{D}^{p}$} & 
        \multicolumn{1}{l}{Training data set of the private model.} \\
        
        \multicolumn{1}{l}{$\mathcal{D}^{a}$} & 
        \multicolumn{1}{l}{Training data set of the attack model.} \\
        
        \midrule
        
        \multicolumn{1}{l}{$\mathbf{M_{s_{1}}}, ..., \mathbf{M_{s_{m}}}$} & 
        \multicolumn{1}{l}{Shadow models.} \\
        
        \multicolumn{1}{l}{$\mathbf{M_{a}}$} & 
        \multicolumn{1}{l}{Attack model.} \\
        
        \midrule
        \multicolumn{1}{l}{$\mathbf{x, y}$} & 
        \multicolumn{1}{l}{Speech data and its emotion label.} \\

        \multicolumn{1}{l}{$\mathbf{z}$} & 
        \multicolumn{1}{l}{Sensitive attribute label.} \\
        
        \multicolumn{1}{l}{$t, k$} & 
        \multicolumn{1}{l}{Global epoch and client index in FL.} \\
        
        \multicolumn{1}{l}{$U$} & 
        \multicolumn{1}{l}{Total number of clients.} \\
        
        \multicolumn{1}{l}{$\mathbf{\theta}^{t}$} & 
        \multicolumn{1}{l}{Global model parameters at $t^\text{th}$ epoch.} \\

        \multicolumn{1}{l}{$\mathbf{\theta}_{k}^{t}$} & 
        \multicolumn{1}{l}{Model updates of $k^\text{th}$ client at $t^\text{th}$ epoch.} \\
        
        \multicolumn{1}{l}{$q$} & 
        \multicolumn{1}{l}{Client sample ratio for each training epoch.} \\
        
        \multicolumn{1}{l}{$T$} & 
        \multicolumn{1}{l}{Total number of global training epoch.} \\
        
        \multicolumn{1}{l}{$C$} & 
        \multicolumn{1}{l}{Norm clipping threshold.} \\
        
        \multicolumn{1}{l}{$n$} & 
        \multicolumn{1}{l}{Number of leaked model updates in FL.} \\

        \bottomrule
    \end{tabular}
    \label{tab:notation}
    \vspace{-1mm}
\end{table}

\vspace{-1mm}
\subsection{Attack Framework}
\label{sec:framework}
We use an attack framework similar to membership inference attack~\cite{shokri2017membership}. Below is a summary of the attack framework, and a more detailed description can be found in \cite{feng2021attribute}.

\vspace{0.5mm}
\noindent \textbf{1. Shadow FL training:} The adversary first trains several shadow SER models $\mathbf{M_{s_{1}}}, \mathbf{M_{s_{2}}}, ..., \mathbf{M_{s_{m}}}$ to mimic the private training on $\mathcal{D}^{p}$. The adversary trains each shadow model using different folds of training data. The data sets for training these shadow models can come from public data sets with similar distribution to $\mathcal{D}^{p}$. We want to underscore that the public data sets used to train the shadow models and private training data set $\mathcal{D}^{p}$ are mutually exclusive in this attack framework. Here, we assume the attack is a white-box attack, where the attacker knows the model architecture and hyper-parameters like batch size and learning rate. Therefore, shadow models have the same model architecture as the private model and have the same training hyper-parameters used in the private training. 

\vspace{0.5mm}

\noindent \textbf{2. Attack data set:} We collect the global model $\mathbf{\theta}$ and trained local model $\mathbf{\theta}_{k}$ of $k^\text{th}$ client at each epoch while training  $\mathbf{M_{s_{1}}}, \mathbf{M_{s_{2}}}, ..., \mathbf{M_{s_{m}}}$ as the attack training data set $\mathcal{D}^{a}$. Here, we further define the \textbf{pseudo gradients} $\mathbf{g'}_{k}^{t}$ as the training input of the attacker model. Given $t$ (number of local training updates) and $\eta$ (learning rate), we can write $\mathbf{g'}_{k}^{t}$ as follows:

\vspace{-0.75mm}
\begin{equation}
    \mathbf{g'}_{k}^{t}  = \frac{1}{t\eta} (\mathbf{\theta^{t}} - \mathbf{\theta^{t}_{k}})
    \label{equ:psuedo_gradient}
\end{equation}

\noindent \textbf{3. Attack model training:} In this paper, the attacker model takes $\mathbf{g'}_{k}^{t}$ as the input to infer $z_{k}$ of the $k^\text{th}$ client. Suppose $\mathbf{\nabla W_{i}}$ and $\mathbf{\nabla b_{i}}$ are the weight updates and the bias updates in $\mathbf{g'}$ corresponding to the $i^\text{th}$ layer in SER training, respectively. Each layer's weight update is first fed into a three-layer CNN feature extractor to compute the hidden representation. We then flatten the output from the CNN module and concatenate it with the layer's bias updates. We then pass this combined representation to a multi-layer perceptron~(MLP) classifier to predict gender. In this work, we focus on using the $\mathbf{\nabla W_{1}}$ and $\mathbf{\nabla b_{1}}$ based on our observation that most information leakage in this application comes from the first layer's training updates \cite{feng2021attribute}.

\vspace{-1.5mm}
\subsection{User-Level Differential Privacy}
The idea of LDP is to perturb the local data using mechanism $\mathcal{M}$ such that the data perturbation is guaranteed to protect from inference attacks given parameters $\epsilon$ and $\delta$. Here, $\epsilon>0$ sets the bound of all outputs on neighboring data sets $\mathcal{D}$ and $\mathcal{D'}$, which differ by one sample, in a database. $\delta\in[0, 1)$ indicates the probability that the ratio of the probabilities for two adjacent data sets $\mathcal{D}$ and $\mathcal{D'}$ cannot be bounded by $\epsilon$. Given a fixed $\delta$, a lower $\epsilon$ represents stronger privacy protection \cite{dwork2006calibrating}. More formally, LDP can be defined as follows:

\vspace{-1.5mm}
\begin{definition}[$(\epsilon, \delta)$-LDP]
    A random mechanism $\mathcal{M}$ satisfies $(\epsilon, \delta)$-LDP, where $\epsilon>0$ and $\delta\in[0, 1)$, if and only if for any two adjacent data sets $\mathcal{D}$ and $\mathcal{D'}$ in universe $\mathcal{X}$, we have:
    
    \vspace{-1.5mm}
    \begin{equation}
        Pr(\mathcal{M}(\mathcal{D})) \leq e^{\epsilon} Pr(\mathcal{M}(\mathcal{D'})) + \delta
    \end{equation}
\end{definition}
\vspace{-1.5mm}

In this paper, we follow the work in \cite{wei2021user} and select Gaussian mechanism using $L_{2}$ norm sensitivity as $\mathcal{M}$. In this setup, we perturb an output $s(x)$ by adding Gaussian noise with zero-mean and variance $\sigma^2\mathbf{I}$ for a given $s(\cdot)$ as shown below:

\vspace{-2mm}
\begin{equation}
    \mathcal{M}(x) = s(x) + \mathcal{N}(0, \sigma^2\mathbf{I})
\end{equation}
\vspace{-3mm}

In the FL setup, the model update function $\ell(\mathcal{D}^{p}, \theta)$ becomes a natural choice for the sample function in the LDP. Formally, the sensitivity is defined as the upper bound for the noise perturbation given by $\sigma$ that satisfies $(\epsilon, \delta)$-LDP. Given two adjacent data sets $\mathcal{D}_{k}^{p}$ and $\mathcal{D'}_{k}^{p}$ and the gradients $g(\mathcal{D}_{k}^{p})=\ell(\mathcal{D}_{k}^{p}, \theta^{t})$ in the $k^\text{th}$ client and $t^\text{th}$ global epoch, the max sensitivity associated with this process is as follows:

\vspace{-1.25mm}
\begin{equation}
    \nabla \ell = \max_{\mathcal{D}_{k}^{p}, \mathcal{D'}_{k}^{p} \in \mathcal{X}} ||g(\mathcal{D}_{k}^{p})-g(\mathcal{D'}_{k}^{p})||_{2} 
\end{equation}
\vspace{-1.25mm}

More specifically, the norm clipping technique in deep learning is frequently used to bound the sensitivity function above \cite{abadi2016deep}. Given the norm clipping threshold $\mathcal{C}$, we can bound the sensitivity as $\nabla \ell \leq \frac{2 \eta \mathcal{C}}{|\mathcal{D}_{k}^{p}|}$. Furthermore, given total training epoch $T$, the number of clients participating in a global epoch $K$, the client sample ratio $q=\frac{K}{U}$, $\epsilon_{k}$, and fixed $\delta_{k}$, the following inequality can be derived as shown in \cite{abadi2016deep} and \cite{wei2021user}:

\vspace{-1.25mm}
\begin{equation}
    \ln{\frac{1}{\delta_{k}}}<\frac{\epsilon_{k}^2 \sigma_{k}^2}{2Tq\nabla \ell^2}
\end{equation}
\vspace{-1.25mm}

Thus, we can determine $\sigma_{k}$ of the Gaussian noise that satisfies $(\epsilon_{k}, \delta_{k})$-LDP for the $k^\text{th}$ client using the equation below:

\vspace{-1.25mm}
\begin{equation}
    \sigma_{k} = \frac{\nabla \ell \sqrt{2qT\ln{(1/\delta_{k})}}}{\epsilon_{k}}
    \label{equ:dp_noise}
\end{equation}
\vspace{-1.25mm}

So unlike the normal FL process, where the local client directly uploads the updated model parameters for aggregation, the UDP framework locally adds Gaussian noise with zero mean and variance $\sigma_{k}$ to $\theta^{t+1}_{k}$ before sending it to the central server. Algorithm \ref{alg:fed-udp} shows the federated learning with UDP. Additionally, for a given $\epsilon_{k}$, a larger $T$ in the entire training process leads to lower privacy guarantees because the adversary may access more observations of model updates \cite{wei2021user}. This decrease in privacy protection can be related to the composition property associated with DP derived in \cite{dwork2010boosting, dwork2006calibrating}:

\begin{algorithm}
    \caption{User-level DP (UDP)}
    \begin{algorithmic}[1]
        \State{\textbf{Initialize: }}{$\mathbf{\theta^{0}}$, $\mathbf{c^{0}}, q, T, C,$ LDP parameters $(\epsilon_{i}, \delta_{i})$ for every client}
        
        \For{Each round $t=0,...,T-1$}
            \State{Sample clients $\mathcal{S}\in\{1, 2, ..., U\}$}
            
            \For{Each client $k \in \mathcal{S}$ in parallel}
                \State $g_{k}^{t} (D^{p}_{k}) \gets \ell(\mathcal{D}_{k}^{p}, \theta^{t})$
                
                \State $g_{k}^{t} (D^{p}_{k}) \gets g_{k}^{t} (D^{p}_{k}) / \max(1, \frac{||g_{k}^{t} (D^{p}_{k})||_{2}}{C})$

                \State $\mathbf{\theta_{k}^{t+1}} \gets \mathbf{\theta^{t}} - \eta g_{k}^{t} (D^{p}_{k})$
                
                \State $\sigma_{k} \gets \frac{\nabla \ell \sqrt{2qT\ln{(1/\delta_{k})}}}{\epsilon_{k}}$

                \State $\mathbf{\theta_{k}^{t+1}} \gets \mathbf{\theta_{k}^{t+1}} + \mathcal{N}(0, \sigma_{k}\mathbf{I})$
                
            \EndFor{\textbf{end}}
            
            \State $\mathbf{\theta^{t+1}} \gets \frac{1}{|\mathcal{S}|} \sum_{k\in \mathcal{S}} \mathbf{\theta_{k}^{t+1}}$
            
        \EndFor{\textbf{end}}

    \end{algorithmic}
    \label{alg:fed-udp}
\end{algorithm}

\vspace{-3.25mm}
\begin{theorem}
    For any $\epsilon>0$ and $\delta\in[0, 1)$, the class of ($\epsilon, \delta$)-DP mechanisms satisfy ($k\epsilon, k\delta$)-DP under k-fold composition.
\end{theorem}
\vspace{-1.25mm}

Therefore, we hypothesize that the attack performance increases with more model updates leaked. Finally, we test the attack performance by varying the number of leaked observations, $n$, of a client to empirically validate this behavior.

\vspace{-1mm}

\section{SER Data Sets}
\label{sec:data}

In this work, we use three corpora widely used in SER, including in our previous attacker work, to evaluate the DP performance. Readers can reference the label distribution of the data set in \cite{feng2021attribute}.

\vspace{1mm}

\noindent \textbf{1. The IEMOCAP} database \cite{busso2008iemocap} contains audio and visual data of acted human interactions with categorical emotions. The corpus has five recorded sessions from ten subjects (five male and five female) in scripted and improvised conditions. Speakers follow a fixed script in the scripted condition and perform spontaneous interactions in the improvised condition. Similar to \cite{zhang2018attention} and our previous work \cite{feng2021attribute}, we only use the data from the improvised condition. We decided to use the four most frequently occurring emotion labels (neutral, sad, happiness, and anger) for training the SER model as suggested in \cite{zhang2018attention}.

\vspace{1mm}

\noindent \textbf{2. The CREMA-D} \cite{cao2014crema} corpus has 7,442 speech recordings that simulate different emotional expressions. The whole database is collected from 91 actors (48 male and 43 female).

\vspace{1mm}

\noindent \textbf{3. The MSP-Improv} \cite{busso2016msp} corpus consists of human interactions with naturalistic emotions captured from improvised scenarios. The whole data set is from 12 participants (six male and six female). Like the IEMOCAP data set, we only select data recorded in the improvised condition. 

\vspace{-1mm}

\section{Experiments}

\subsection{Data Preprocessing}

We follow the data preprocessing from our previous work \cite{feng2021attribute}, where we extract the EmoBase feature set and the autoregressive predictive coding (APC) \cite{chung2019unsupervised} feature set of each utterance using the OpenSMILE toolkit \cite{eyben2010opensmile} and SUPERB (Speech Processing Universal PERformance Benchmark) \cite{yang21c_interspeech}, respectively. 
We present results on one knowledge-based feature set (EmoBase) and one deep-learning-based feature set (APC). Due to space constraints in the paper, we present the results using other deep-learning-based speech features in our GitHub repository mentioned earlier. We apply z-normalization to the speech features of each speaker. For the IEMOCAP and the MSP-Improv data set, we divide each speaker's data into $10$ \textit{shards} of equal size to create more clients for the FL training. We leave 20\% of speakers as the test data and repeat the experiments five times with test folds of different speakers.

\begin{table*}

    \centering
    \caption{Prediction results of the SER model and the pre-trained attacker model on private data set $\mathcal{D}^{p}$. The \% unweighted average recall (UAR) scores of the SER task and the adversary task (gender inference) on each data set are reported. $\epsilon$ indicates the privacy level set in the UDP algorithm, and a smaller $\epsilon$ represents a stronger privacy guarantee. }
    
    \vspace{-1.25mm}
    
    \begin{tabular}{>{\centering\arraybackslash}p{16mm}>{\centering\arraybackslash}p{16mm}>{\centering\arraybackslash}p{8.5mm}>{\centering\arraybackslash}p{8.5mm}>{\centering\arraybackslash}p{8.5mm}>{\centering\arraybackslash}p{8.5mm}>{\centering\arraybackslash}p{8.5mm}>{\centering\arraybackslash}p{8.5mm}>{\centering\arraybackslash}p{8.5mm}>{\centering\arraybackslash}p{8.5mm}>{\centering\arraybackslash}p{8.5mm}>{\centering\arraybackslash}p{9mm}}
        
        \toprule
        
        \multirow{2}{*}{$\mathcal{D}^p$} & 
        \multirow{2}{*}{\textbf{Feature}} &
        \multicolumn{5}{c}{\textbf{SER Performance(\% UAR)}} &
        \multicolumn{5}{c}{\textbf{Attacker Performance(\% UAR)}} \rule{0pt}{1.65ex} \\
        
        \cmidrule(lr){3-7} \cmidrule(lr){8-12}
        
        & &
        $\mathbf{\epsilon=\infty}$ &
        $\mathbf{\epsilon=50}$ &
        $\mathbf{\epsilon=25}$ &
        $\mathbf{\epsilon=10}$ &
        $\mathbf{\epsilon=5}$ &
        $\mathbf{\epsilon=\infty}$ &
        $\mathbf{\epsilon=50}$ &
        $\mathbf{\epsilon=25}$ &
        $\mathbf{\epsilon=10}$ &
        $\mathbf{\epsilon=5}$ \rule{0pt}{1.65ex} \\
        \cmidrule(lr){1-1} \cmidrule(lr){2-2}
        \cmidrule(lr){3-7} \cmidrule(lr){8-12}
        
        \multirow{2}{*}{\textbf{IEMOCAP}} & 
        \textbf{EmoBase} & 
        61.6 & 
        60.7 & 
        58.5 & 
        59.6 & 
        54.5 & 
        82.5 & 
        51.7 & 
        50.5 & 
        50.2 & 
        50.0 
        \rule{0pt}{1.65ex} \\
        
        & 
        \textbf{APC} & 
        63.4 & 
        61.5 & 
        60.6 & 
        60.0 & 
        54.6 & 
        90.7 & 
        60.4 & 
        53.2 & 
        51.9 & 
        48.9 
        \rule{0pt}{1.95ex} \\
        
        \hline
        \multirow{2}{*}{\textbf{CREMA-D}} & 
        \textbf{EmoBase} & 
        66.3 & 
        64.1 & 
        64.5 & 
        63.0 & 
        61.6 & 
        80.1 & 
        69.3 & 
        58.6 & 
        52.2 & 
        50.1 
        \rule{0pt}{1.95ex} \\
        
        & 
        \textbf{APC} & 
        66.2 & 
        66.0 & 
        65.2 & 
        64.8 & 
        63.2 & 
        78.4 & 
        64.0 & 
        53.9 & 
        50.0 & 
        50.0 
        \rule{0pt}{1.95ex} \\
        
        \hline
        \multirow{2}{*}{\textbf{MSP-Improv}} & 
        \textbf{EmoBase} & 
        47.0 & 
        47.1 & 
        46.1 & 
        46.2 & 
        43.5 & 
        89.1 & 
        62.2 & 
        53.8 & 
        50.6 & 
        49.3 
        \rule{0pt}{1.95ex} \\
        
        & 
        \textbf{APC} & 
        51.1 & 
        51.3 & 
        50.7 & 
        48.7 & 
        45.6 & 
        93.2 & 
        59.5 & 
        52.1 & 
        49.8 & 
        50.0 
        \rule{0pt}{1.95ex} \\

        \hline

    \end{tabular}
    
    \label{tab:fl_result}
    \vspace{-2mm}
\end{table*}

\begin{figure*}[ht] {
    \centering
    
    \begin{tikzpicture}
        
        \node[draw=none, align=center] at (0.34\linewidth, 4.9){\textbf{Attack performance}};
        
        \node[draw=none,fill=none] at (0, 3.1){\includegraphics[width=0.33\linewidth]{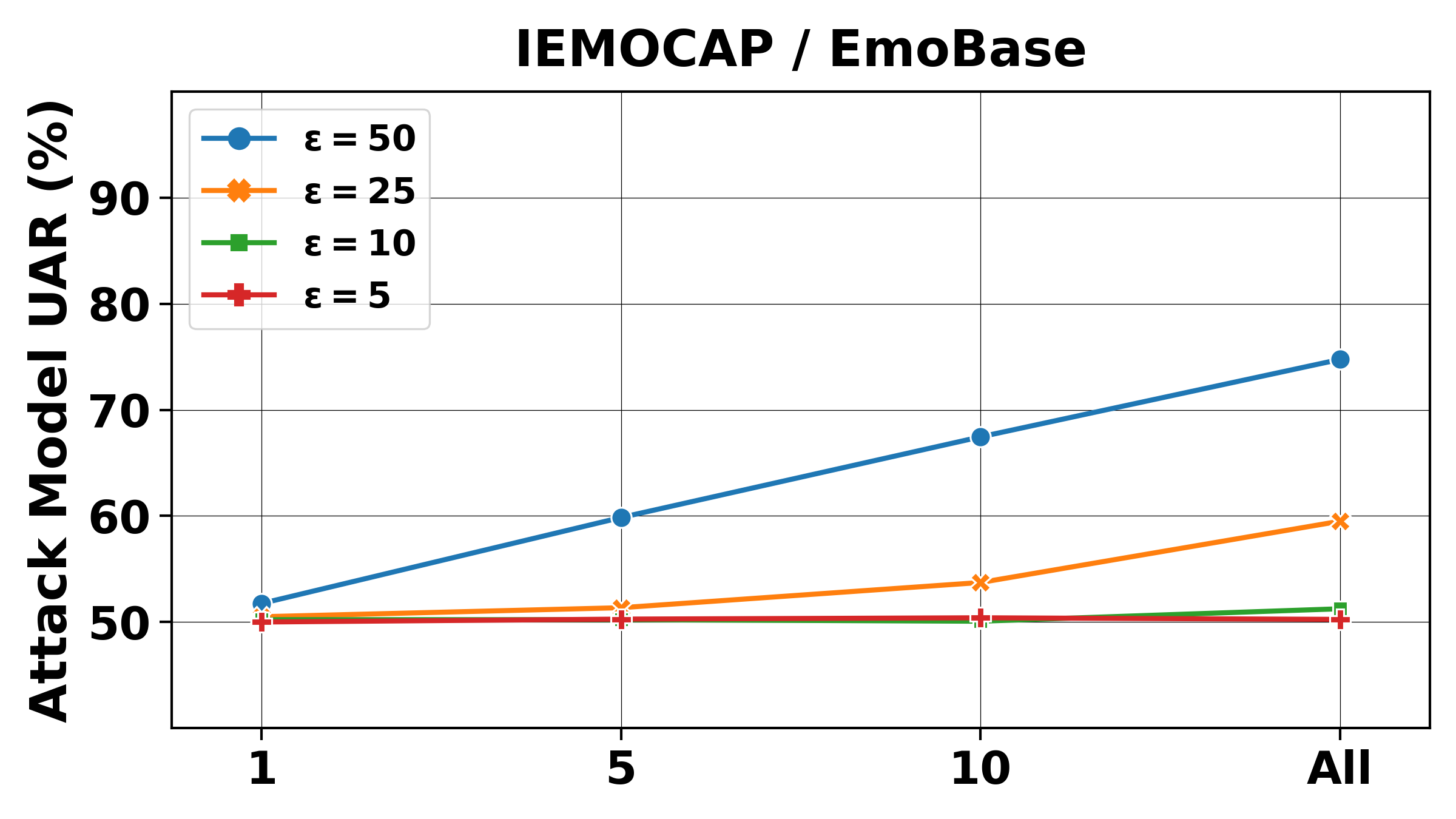}};
        
        \node[draw=none,fill=none] at (0.33\linewidth, 3.1){\includegraphics[width=0.33\linewidth]{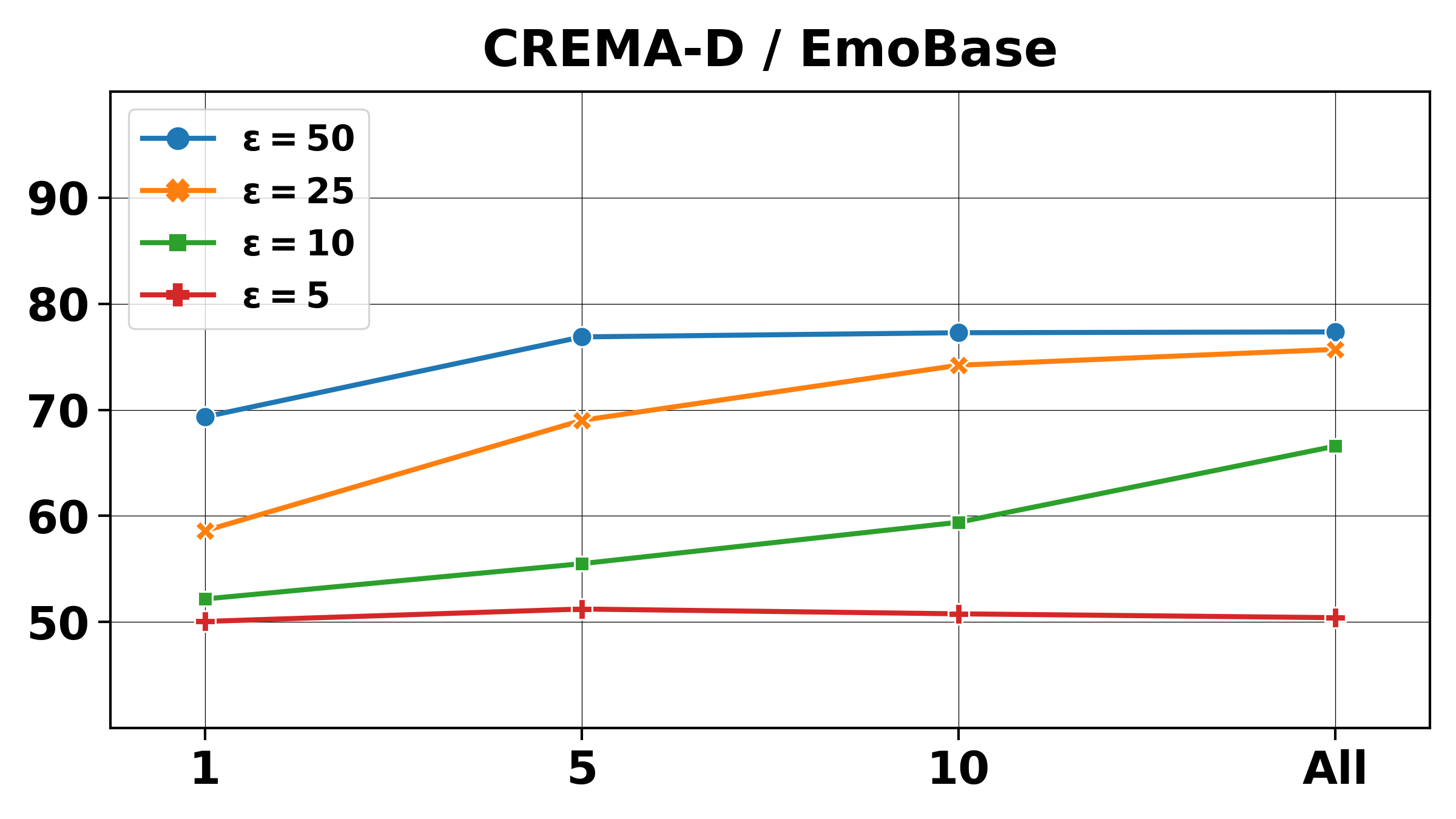}};
        
        \node[draw=none,fill=none] at (0.66\linewidth, 3.1){\includegraphics[width=0.33\linewidth]{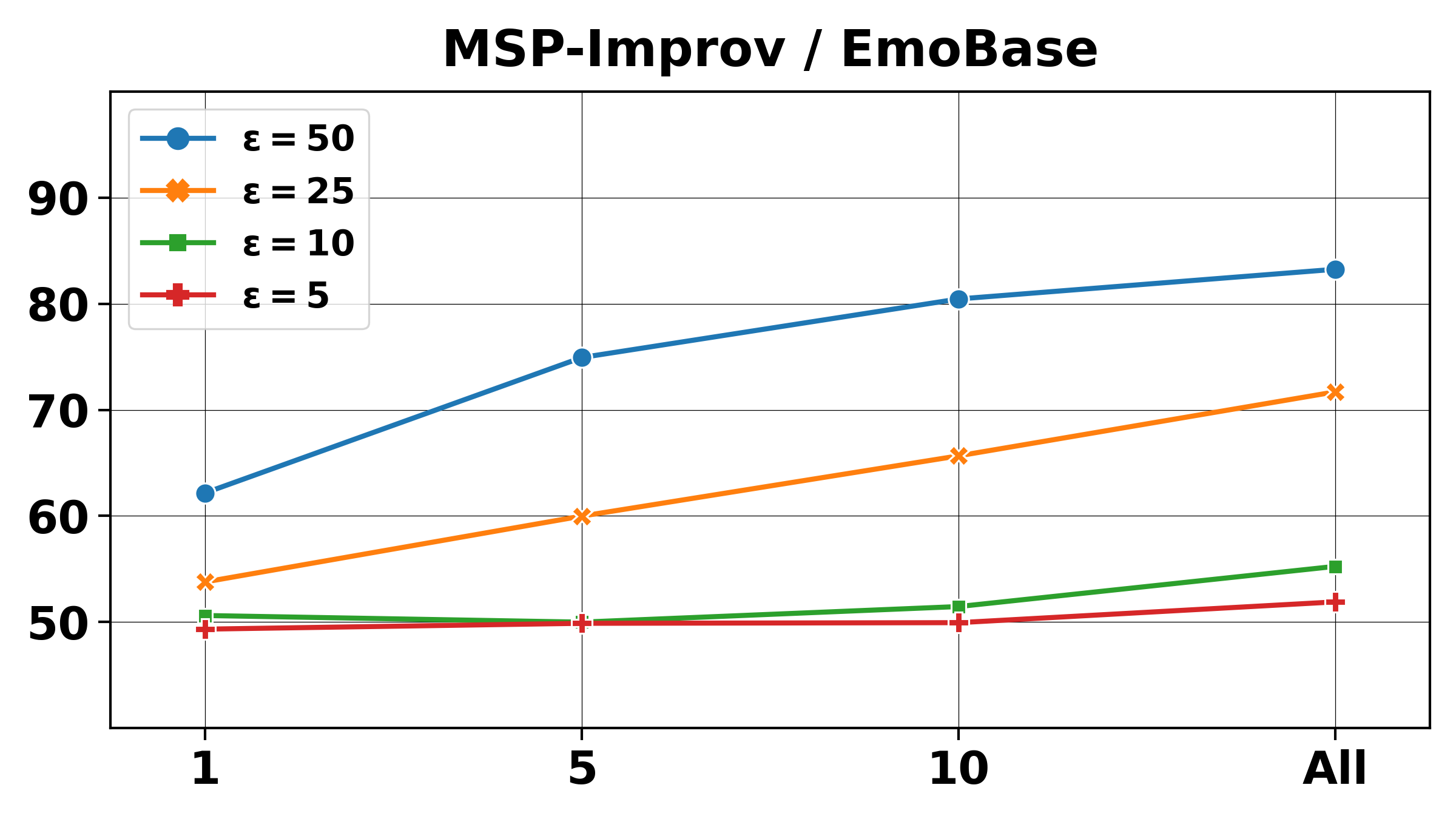}};
        
        \node[draw=none,fill=none] at (0, 0){\includegraphics[width=0.33\linewidth]{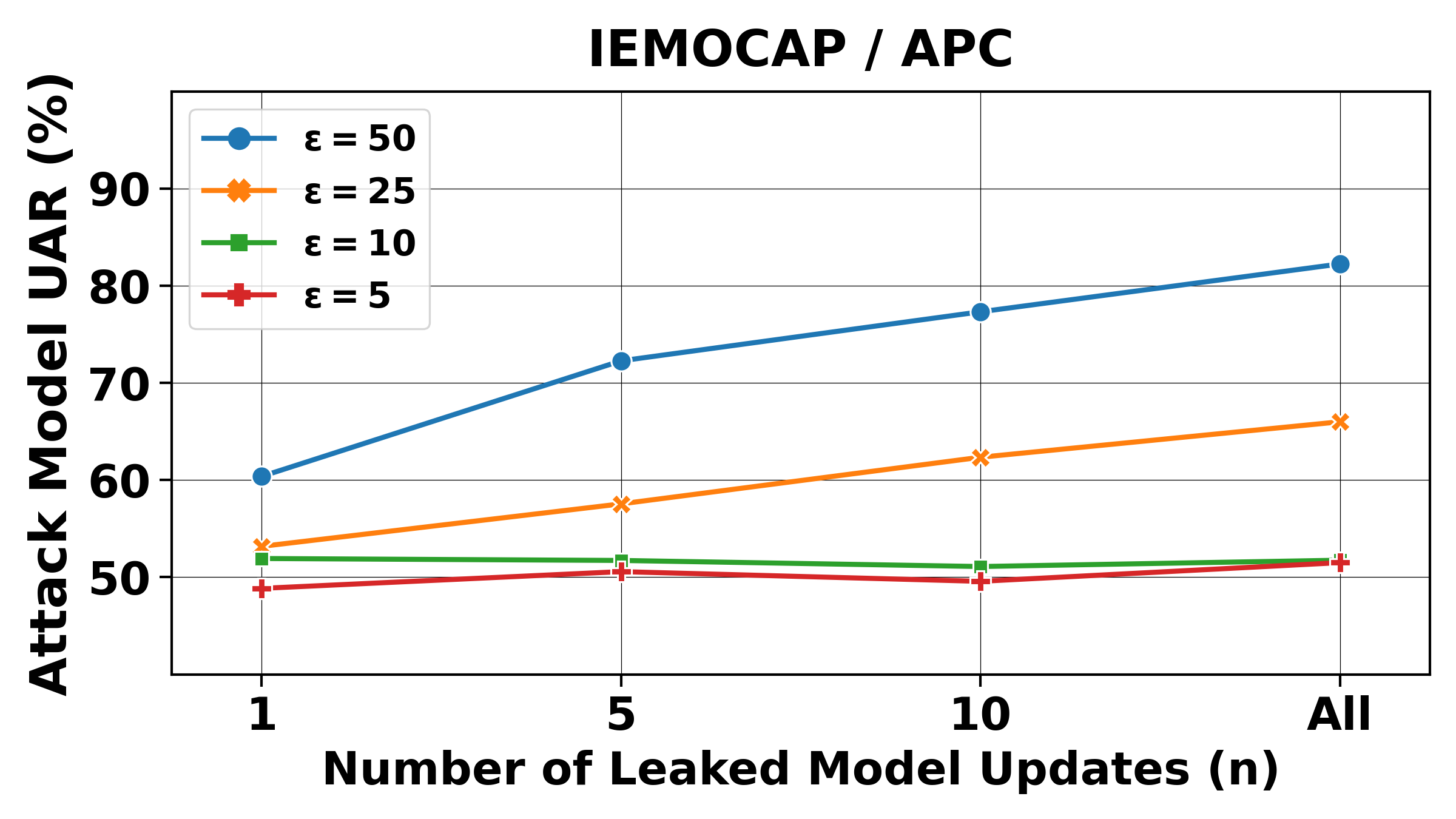}};
        
        \node[draw=none,fill=none] at (0.33\linewidth, 0){\includegraphics[width=0.33\linewidth]{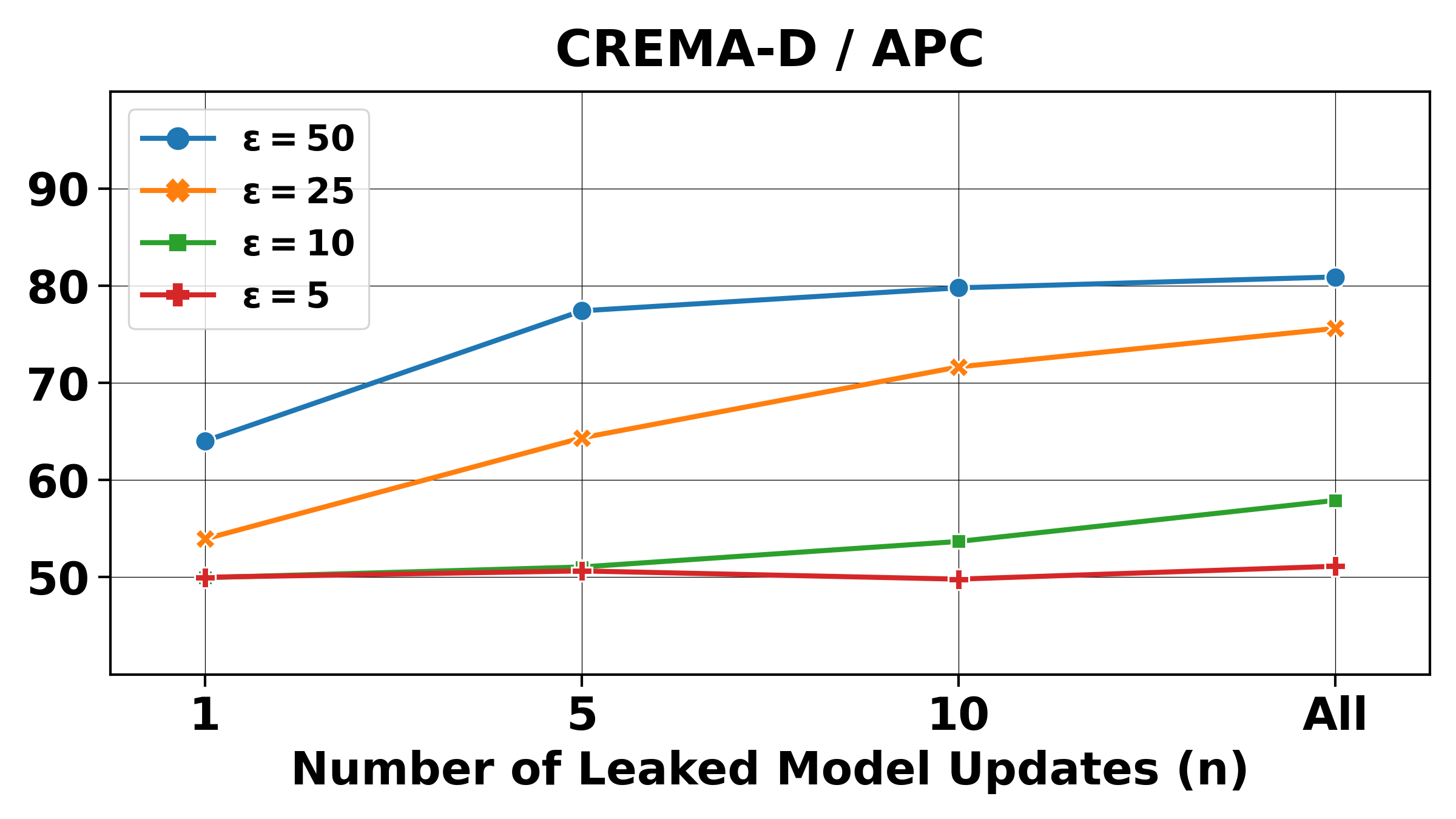}};
        
        \node[draw=none,fill=none] at (0.66\linewidth, 0){\includegraphics[width=0.33\linewidth]{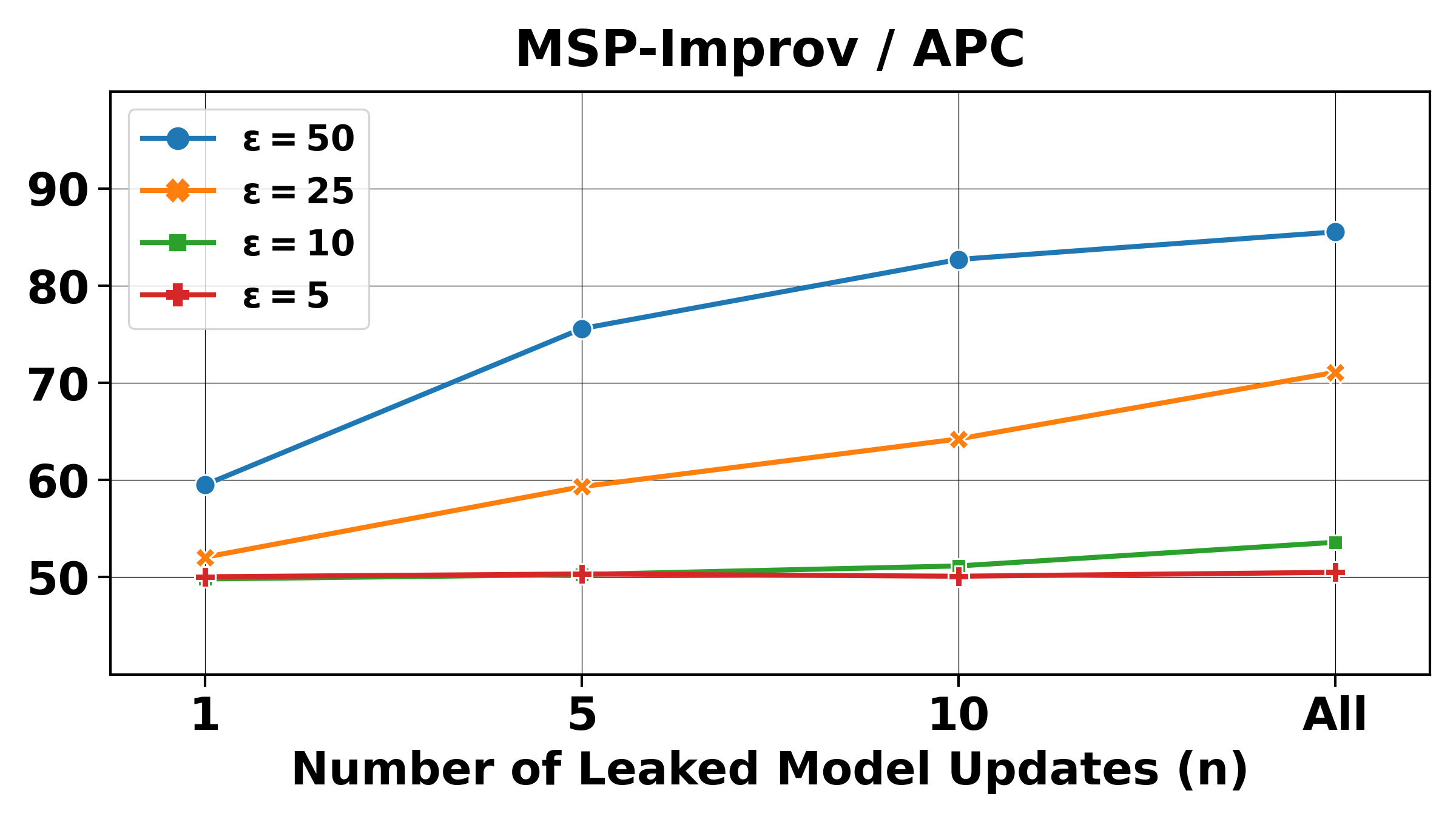}};
        
    \end{tikzpicture}
    
    \vspace{-2.5mm}
    \caption{The figure shows the prediction results of the attribute inference task at different privacy levels ($\epsilon$) and different number of leaked model updates.
    We denote the data set and feature set by the notation $data\ set/feature\ set$.} 
    \label{fig:attack_results}

    \vspace{-3.75mm}
    
} \end{figure*}

\vspace{-1.5mm}
\subsection{Data setup}

Similar to \cite{feng2021attribute}, we simulate the experiments using different private training data sets. For instance, when the IEMOCAP data set is the private training data set $\mathcal{D}^{p}$, the MSP-Improv data set and CREMA-D data set are combined to train shadow models $\mathbf{M_{s_{1}}, ..., M_{s_{m}}}$. Next, we train the attack model $\mathbf{M_{a}}$ using the model updates generated while training $\mathbf{M_{s_{1}}, ..., M_{s_{m}}}$. Finally, we evaluate the performance of $\mathbf{M_{a}}$ using the model updates generated in the FL that uses IEMOCAP data set as $\mathcal{D}^{p}$. Similarly, we repeat the same experiments with the MSP-Improv data set and the CREMA-D data set as $\mathcal{D}^{p}$.

\vspace{-1.5mm}

\subsection{Model and Evaluation Details}

We use an MLP for the SER model architecture. The model consists of 2 dense layers with hidden layer dimensions of \{256, 128\}. We choose ReLU as the activation function and the dropout rate as 0.2. We implement the FedAvg algorithm in training the SER model. Only $q=10\%$ of the clients participate in each global round. 80\% of the data at a client is reserved for local training, and the remaining 20\% is used for validation. We set the local training batch size as 20, the $\eta$ as 0.0005, the local training epoch as 1, and the $T$ as 200. We set the norm clipping threshold $C=0.25$ and $\delta_{k}=0.5$ for every client. We evaluate the attacker performance under several privacy budget values $\epsilon_{k} \in [5, 10, 25, 50]$. We use the pre-trained attacker model from our previous work, and details of the attacker model training are in \cite{feng2021attribute}. We randomly pick a client's $n$ model updates (generated in FL) and predict its gender label using the aggregated model updates. As we mentioned in section \ref{sec:framework}, we only use the model updates from the first layer as the input for the inference task. We repeat this ten times for each client and aggregate predictions from all clients to report the final results. We empirically test $n \in [1, 5, 10, all]$, where \textit{all} refers to the scenario where all the updates available from a client are available to the attacker.

\vspace{-1mm}
\section{Results and Discussion}

\subsection{SER Performance}
The SER results of UDP-based FL at different privacy levels are shown in \autoref{tab:fl_result}. $\epsilon=\infty$ indicates the case of FL without adding UDP. In this work, we report the unweighted average recall (UAR) scores of the SER predictions over the emotion classes. Overall, the SER model performs best in the CREMA-D data set. Across the different datasets and feature sets, we observe that the SER performance decreases by about 1-2\% when applying UDP with $\epsilon=50$ and $\epsilon=25$. Moreover, the UAR decreases by around 3-4\% when $\epsilon$ reduces to 10. Finally, the SER performance drops significantly when $\epsilon=5$ in the UDP. These observations comply with the expected output of UDP, where a relatively larger $\epsilon$ is associated with smaller noises added to the model parameters and thus does not substantially impact the performance of the primary application. To quantify the amount of noise added to the weight parameters, we calculate the weight parameters' signal-to-noise ratio (SNR) at different privacy levels. We find that the SNR is in the range of 14.11 dB to 20.65 dB when $\epsilon=25$, which suggests that the SER model performance decreases substantially when the energy of the shared weight parameters is less than 25 times the energy of the noise.

\vspace{-1.25mm}
\subsection{Attacker Performance \texorpdfstring{($\mathbf{n=1}$)}{}}
The attacker results of FL with UDP at different $\epsilon$ are shown in \autoref{tab:fl_result}. Similar to the SER performance, we evaluate the attacker using the UAR scores of gender predictions. The table shows that the pre-trained attack model can predict gender with a UAR above 75\% in all conditions when no perturbation is added ($\epsilon=\infty$). However, we find that the gender predictions from the attacker model drop intensely even when applying the UDP using $\epsilon=50$ (small perturbation $\sigma_k$). As we reduce $\epsilon$ from 50 to 25 (hence gradually increasing the perturbation), the gender prediction results drop close to the random guess. These results indicate that UDP can effectively mitigate the attribute inference attack without sacrificing much utility of the SER model when the attacker has only access to a single model update from a client.

\vspace{-1.25mm}
\subsection{Attacker Performance \texorpdfstring{($\mathbf{n>1}$)}{}}

\autoref{fig:attack_results} shows the results of the attack performance on FL with access to a varying number of model updates, $n$ and at different privacy levels, $\epsilon$ for the UDP algorithm. The results show that the attack performance in gender prediction improves by an enormous margin with more model updates of a client leaked to the adversary when $\epsilon=50$ and $\epsilon=25$. For example, the UAR of the gender prediction is 82.3\% when the adversary has access to all model updates of a client and $\epsilon=50$ in UDP, which shows that the attacker can infer the gender to a reasonable extent with access to all model updates from a client. However, at $\epsilon=10$ and $\epsilon=5$, the attack performance does not increase much, even with more access to model updates. These results suggest that when the attacker can observe multiple model updates from the UDP-based FL process, the attribute inference attack is achievable with some degradation in the SER performance by applying a small $\epsilon$ in UDP.

\vspace{-1.5mm}
\section{Conclusions}

We evaluated the attribute inference attack of the SER task within FL settings with a user-level DP algorithm. The UDP algorithm used in this paper satisfies the requirement of LDP with privacy parameters $\epsilon$ and $\delta$. We discover that the UDP algorithm can effectively mitigate the attribute inference attack when the adversary can only access one model update from a client. This defense provides promising results even with a relatively larger $\epsilon$ at 50 (weaker privacy guarantee). However, as the number of leaked model updates increases, the adversary can infer the gender label with an adequate UAR when $\epsilon$ are 50 and 25. Since the current adversary trains the attack model using the model updates generated from only two public SER data sets, the attacker can potentially improve the performance of the attack model by including more public SER data sets. Consequently, this may make UDP less effective against the current attribute inference attack framework. Therefore, in future works, we aim to explore adversarial training, which targets to protect specific attributes in the defense.

\bibliographystyle{IEEEtran}

\bibliography{mybib}

\begin{thebibliography}{10}
\providecommand{\url}[1]{#1}
\csname url@samestyle\endcsname
\providecommand{\newblock}{\relax}
\providecommand{\bibinfo}[2]{#2}
\providecommand{\BIBentrySTDinterwordspacing}{\spaceskip=0pt\relax}
\providecommand{\BIBentryALTinterwordstretchfactor}{4}
\providecommand{\BIBentryALTinterwordspacing}{\spaceskip=\fontdimen2\font plus
\BIBentryALTinterwordstretchfactor\fontdimen3\font minus
  \fontdimen4\font\relax}
\providecommand{\BIBforeignlanguage}[2]{{%
\expandafter\ifx\csname l@#1\endcsname\relax
\typeout{** WARNING: IEEEtran.bst: No hyphenation pattern has been}%
\typeout{** loaded for the language `#1'. Using the pattern for}%
\typeout{** the default language instead.}%
\else
\language=\csname l@#1\endcsname
\fi
#2}}
\providecommand{\BIBdecl}{\relax}
\BIBdecl

\bibitem{lee2020study}
M.-C. Lee, S.-Y. Chiang, S.-C. Yeh, and T.-F. Wen, ``Study on emotion
  recognition and companion chatbot using deep neural network,''
  \emph{Multimedia Tools and Applications}, vol.~79, no.~27, pp.
  19\,629--19\,657, 2020.

\bibitem{ramakrishnan2013speech}
S.~Ramakrishnan and I.~M. El~Emary, ``Speech emotion recognition approaches in
  human computer interaction,'' \emph{Telecommunication Systems}, vol.~52,
  no.~3, pp. 1467--1478, 2013.

\bibitem{Bone2017SignalProcessingandMachine}
D.~Bone, C.-C. Lee, T.~Chaspari, J.~Gibson, and S.~Narayanan, ``Signal
  processing and machine learning for mental health research and clinical
  applications,'' \emph{IEEE Signal Processing Magazine}, vol.~34, no.~5, pp.
  189--196, September 2017.

\bibitem{li2007speech}
W.~Li, Y.~Zhang, and Y.~Fu, ``Speech emotion recognition in e-learning system
  based on affective computing,'' in \emph{Third International Conference on
  Natural Computation (ICNC 2007)}, vol.~5.\hskip 1em plus 0.5em minus
  0.4em\relax IEEE, 2007, pp. 809--813.

\bibitem{mcmahan2017communication}
B.~McMahan, E.~Moore, D.~Ramage, S.~Hampson, and B.~A. y~Arcas,
  ``Communication-efficient learning of deep networks from decentralized
  data,'' in \emph{Artificial intelligence and statistics}.\hskip 1em plus
  0.5em minus 0.4em\relax PMLR, 2017, pp. 1273--1282.

\bibitem{feng2021attribute}
T.~Feng, H.~Hashemi, R.~Hebbar, M.~Annavaram, and S.~S. Narayanan, ``Attribute
  inference attack of speech emotion recognition in federated learning
  settings,'' \emph{arXiv preprint arXiv:2112.13416}, 2021.

\bibitem{wei2021user}
K.~Wei, J.~Li, M.~Ding, C.~Ma, H.~Su, B.~Zhang, and H.~V. Poor, ``User-level
  privacy-preserving federated learning: Analysis and performance
  optimization,'' \emph{IEEE Transactions on Mobile Computing}, 2021.

\bibitem{melis2019exploiting}
L.~Melis, C.~Song, E.~De~Cristofaro, and V.~Shmatikov, ``Exploiting unintended
  feature leakage in collaborative learning,'' in \emph{2019 IEEE Symposium on
  Security and Privacy (SP)}.\hskip 1em plus 0.5em minus 0.4em\relax IEEE,
  2019, pp. 691--706.

\bibitem{zhu2020deep}
L.~Zhu and S.~Han, ``Deep leakage from gradients,'' in \emph{Federated
  learning}.\hskip 1em plus 0.5em minus 0.4em\relax Springer, 2020, pp. 17--31.

\bibitem{geng2021towards}
J.~Geng, Y.~Mou, F.~Li, Q.~Li, O.~Beyan, S.~Decker, and C.~Rong, ``Towards
  general deep leakage in federated learning,'' \emph{arXiv preprint
  arXiv:2110.09074}, 2021.

\bibitem{wang2019beyond}
Z.~Wang, M.~Song, Z.~Zhang, Y.~Song, Q.~Wang, and H.~Qi, ``Beyond inferring
  class representatives: User-level privacy leakage from federated learning,''
  in \emph{IEEE INFOCOM 2019-IEEE Conference on Computer Communications}.\hskip
  1em plus 0.5em minus 0.4em\relax IEEE, 2019, pp. 2512--2520.

\bibitem{wei2020federated}
K.~Wei, J.~Li, M.~Ding, C.~Ma, H.~H. Yang, F.~Farokhi, S.~Jin, T.~Q. Quek, and
  H.~V. Poor, ``Federated learning with differential privacy: Algorithms and
  performance analysis,'' \emph{IEEE Transactions on Information Forensics and
  Security}, vol.~15, pp. 3454--3469, 2020.

\bibitem{shokri2017membership}
R.~Shokri, M.~Stronati, C.~Song, and V.~Shmatikov, ``Membership inference
  attacks against machine learning models,'' in \emph{2017 IEEE Symposium on
  Security and Privacy (SP)}.\hskip 1em plus 0.5em minus 0.4em\relax IEEE,
  2017, pp. 3--18.

\bibitem{openmoji}
``Openmoji,'' https://openmoji.org/.

\bibitem{dwork2006calibrating}
C.~Dwork, F.~McSherry, K.~Nissim, and A.~Smith, ``Calibrating noise to
  sensitivity in private data analysis,'' in \emph{Theory of cryptography
  conference}.\hskip 1em plus 0.5em minus 0.4em\relax Springer, 2006, pp.
  265--284.

\bibitem{abadi2016deep}
M.~Abadi, A.~Chu, I.~Goodfellow, H.~B. McMahan, I.~Mironov, K.~Talwar, and
  L.~Zhang, ``Deep learning with differential privacy,'' in \emph{Proceedings
  of the 2016 ACM SIGSAC conference on computer and communications security},
  2016, pp. 308--318.

\bibitem{dwork2010boosting}
C.~Dwork, G.~N. Rothblum, and S.~Vadhan, ``Boosting and differential privacy,''
  in \emph{2010 IEEE 51st Annual Symposium on Foundations of Computer
  Science}.\hskip 1em plus 0.5em minus 0.4em\relax IEEE, 2010, pp. 51--60.

\bibitem{busso2008iemocap}
C.~Busso, M.~Bulut, C.-C. Lee, A.~Kazemzadeh, E.~Mower, S.~Kim, J.~N. Chang,
  S.~Lee, and S.~S. Narayanan, ``{IEMOCAP}: Interactive emotional dyadic motion
  capture database,'' \emph{Language resources and evaluation}, vol.~42, no.~4,
  pp. 335--359, 2008.

\bibitem{zhang2018attention}
Y.~Zhang, J.~Du, Z.~Wang, J.~Zhang, and Y.~Tu, ``Attention based fully
  convolutional network for speech emotion recognition,'' in \emph{2018
  Asia-Pacific Signal and Information Processing Association Annual Summit and
  Conference (APSIPA ASC)}.\hskip 1em plus 0.5em minus 0.4em\relax IEEE, 2018,
  pp. 1771--1775.

\bibitem{cao2014crema}
H.~Cao, D.~G. Cooper, M.~K. Keutmann, R.~C. Gur, A.~Nenkova, and R.~Verma,
  ``Crema-d: Crowd-sourced emotional multimodal actors dataset,'' \emph{IEEE
  transactions on affective computing}, vol.~5, no.~4, pp. 377--390, 2014.

\bibitem{busso2016msp}
C.~Busso, S.~Parthasarathy, A.~Burmania, M.~AbdelWahab, N.~Sadoughi, and E.~M.
  Provost, ``Msp-improv: An acted corpus of dyadic interactions to study
  emotion perception,'' \emph{IEEE Transactions on Affective Computing},
  vol.~8, no.~1, pp. 67--80, 2016.

\bibitem{chung2019unsupervised}
Y.-A. Chung, W.-N. Hsu, H.~Tang, and J.~Glass, ``An unsupervised autoregressive
  model for speech representation learning,'' in \emph{Interspeech}, 2019.

\bibitem{eyben2010opensmile}
F.~Eyben, M.~W{\"o}llmer, and B.~Schuller, ``Opensmile: the munich versatile
  and fast open-source audio feature extractor,'' in \emph{Proceedings of the
  18th ACM international conference on Multimedia}, 2010, pp. 1459--1462.

\bibitem{yang21c_interspeech}
S.~wen Yang, P.-H. Chi, Y.-S. Chuang, C.-I.~J. Lai, K.~Lakhotia, Y.~Y. Lin,
  A.~T. Liu, J.~Shi, X.~Chang, G.-T. Lin, T.-H. Huang, W.-C. Tseng, K.~tik Lee,
  D.-R. Liu, Z.~Huang, S.~Dong, S.-W. Li, S.~Watanabe, A.~Mohamed, and
  H.~yi~Lee, ``{SUPERB: Speech Processing Universal PERformance Benchmark},''
  in \emph{Proc. Interspeech 2021}, 2021, pp. 1194--1198.

\end{thebibliography}

\end{document}